# Amazon Product Recommender System

**Mohammad R. Rezaei**

*Department of Biomedical Engineering*

*University of Toronto*

*Toronto, ON M5S 3G9, Canada*

mr.rezaei@mail.utoronto.ca

## 1. Introduction

The number of reviews on Amazon has grown significantly over the years. Customers who made purchases on Amazon provide reviews by rating the product from 1 to 5 stars and sharing a text summary of their experience and opinion of the product [1]. The ratings of a product are averaged to provide an overall product rating. We analyzed what ratings score customers give to a specific product (a music track) in order to build a recommender model for digital music tracks on Amazon. We test various traditional models along with our proposed deep neural network (DNN) architecture to predict the reviews' rating score. The Amazon review dataset contains 200,000 data samples; we train the models on 70% of the dataset and test the performance of the models on the remaining 30% of the dataset.

## 2. Data preprocessing and feature selection

The Amazon review dataset contains 200,000 data samples and each sample has ten different attributes shown in **Figure 1.A**, which described as follows:

- "**reviewerID**" The ID of the user
- "**itemID**" The ID of the item which being reviewed by users
- "**reviewText**" text from the users about the items
- "**summary**" A short summary of the reviews
- "**overall**" The rating score for the items by the users (target values in this problem; see its distribution in **Figure 1.E**)
- "**price**" Price of the items
- "**reviewHash**" a unique identifier for the reviews
- "**unixReviewTime**" Time of the reviews in seconds since 1970
- "**reviewTime**" Plain-text representation of the review times
- "**category**" Category labels of the items being reviewed

We first do simple statistical analysis to understand which of the attributes are valuable to be considered as features for the product recommender model. We count the number of unique values related to each item (see **Figure 1.B**). As can be seen there is some valuable and unique information in the dataset that can be considered as a feature for the recommender. For example, in **Figure 1.C** and **1.D** "category" and "unixReviewTime" have valuable information (they have non-uniformly distributed histogram). Also, **Figure 1.F** and **1.G** show that the reviewer feedback and summary of it have information too. So, we can consider these attributes of data and extract valuable information for the recommender mode from them.

## 2.1 Feature engineering

**A.** To extract features from "reviewText" and "summary" of the reviews we concatenated them to shape a single text and then transform our data into normalized unigrams. We perform the following operations on the raw text for each entry

a) convert all capitalized word to lower
b) Tokenizing the text by continuous sequences of alphabetic characters (remove whitespace, symbols, punctuation, numerals)
c) Clean up the punctuation and do extract stems from each word
d) Finally, we create a fixed-length vector to represent each "reviewText"and "summary" for each review

**B.** We used one-hot encoding for the product category which contains 5 unique values (Pop, Alternative Rock, Jazz, Classical, Dance & Electronic) and considered them as a vector of features.

**C.** We normalized "unixReviewTime" between zero and one and considered it as a feature.

**D**. We also engineered two other features from the data:

- Reviewer History: the number of reviews a reviewer has written.
- Product Popularity: number of reviews that has written for a product.

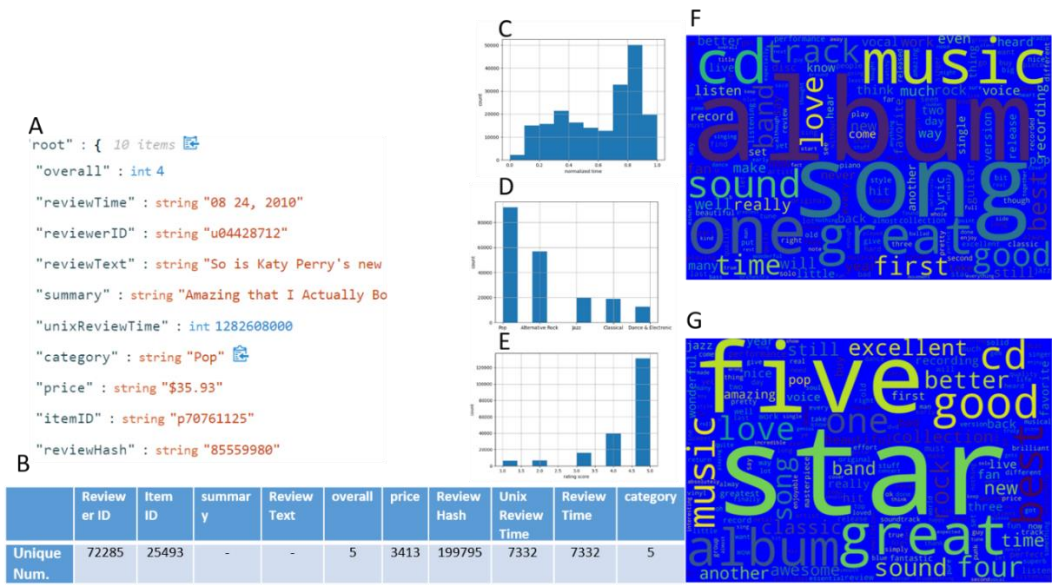

**Figure 1**. A. Data sample from the Amazon review dataset. B Number of unique elements of the data attributes in the whole 200,0000 reviews. C. number of reviews for normalized "unixReviewTime". D. Number of reviews per music them ("category"). E. Number of reviews for rating scores ("overall"). F. frequently of words in "reviewText" for all reviews. .
G. frequently of words in "summary" for all reviews.

### 3. Deep Neural Network (DNN) as product recommender

Based on the high dimensionality of the features and the good performance of DNNs in NLP problems, we suggest a DNN architecture to rate the reviews [2]. As can be seen from **Figure 2**, the model consist of two major parts. The first part, the feature extractor or encoder, extract useful features from the reviews, and the second part, rate estimator, estimate the rate of the review based on the features. Text feature extractor unit consists of a Embedding layer ( turns positive integers (indexes) into dense vectors of fixed size) and followed by a LSTM layer to extract features from the encoded text. In sum, we have 146 features at the end including 100 features from the review text and summary, 5 features of category, 20 features for "itemID", 20 features for "reviewerID", and one for the time of the review.

Finally we concatenate them and use a Dense layer to estimate the rate. We use Adam optimization method [3] with learning rate 0.01 with mean-squared-error (MSE) loss and train the model for 100 epochs.

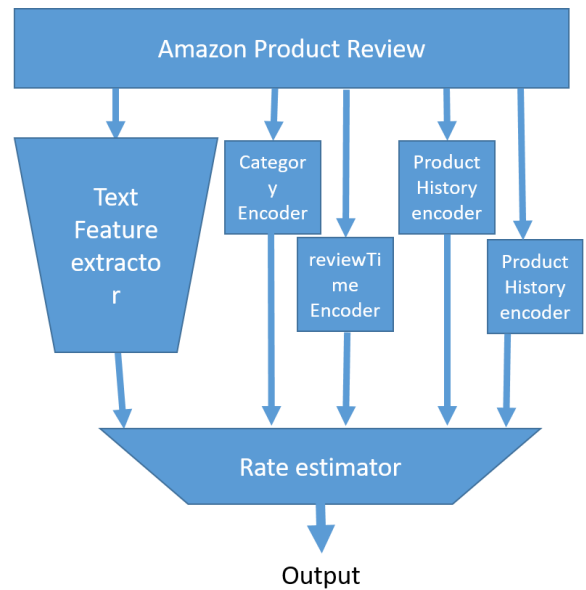

**Figure 2.** Block diagram of the suggested DNN

### 4. Result

Now we compare our DNN with other traditional models like multinomial naive Bayes [4] (which are simple probabilistic models based on applying Bayes theorem with strong independence assumption between features) and logistic regression models [5, 6]. We use MSE as our metric which is suitable for the competition with the competition score.

| Table 1. Summary of the recommender systems performance ||
|---|---|
|  | MSE |
| Multinomial Naive Bayes | 1.29 |
| Logistic Regression | 1.03 |
| DNN | **0.51-0.53** |

As the result in Table 1. shows our DNN model outperforms other methods significantly and shows a promising result to be considered as a recommender system with good performance. Note that even though the RL and MNB models are not as accurate as the DNN is, but they are relatively simple and they need no specific effort to setup them but in the end, they not accurate enough. So they can be considered in problems with no need for high accuracy.

## 5. Discussion

In this project, I faced a couple of challenges to finally reach a good recommender model.

The first one was text preprocessing and encoding: I wasn't familiar with NLP field and it was hard for me to finally find out what approach should I take to preprocess the text data and encode them. So finally I reached to this cheat sheet. 1) eliminate all of the special characters 2) transform the words to lower case words 3) fix the punctuation and eliminate useless words (a, an, is, are…) 4) replace the words their stems 5) tokenize text 6) encoded them in a fixed-length feature map.

Another challenge was the model selection. The number of data is limited to 200,000 and I needed to limit the model's size (number of parameters) small enough to prevent overfitting to the training data and simultaneously maintain a good prediction performance. I test several different architectures and different combinations of layers to reach a small but efficient model. **Figure 3** shows training and validation loss over training session, which indicates no overfitting of the model on training data.

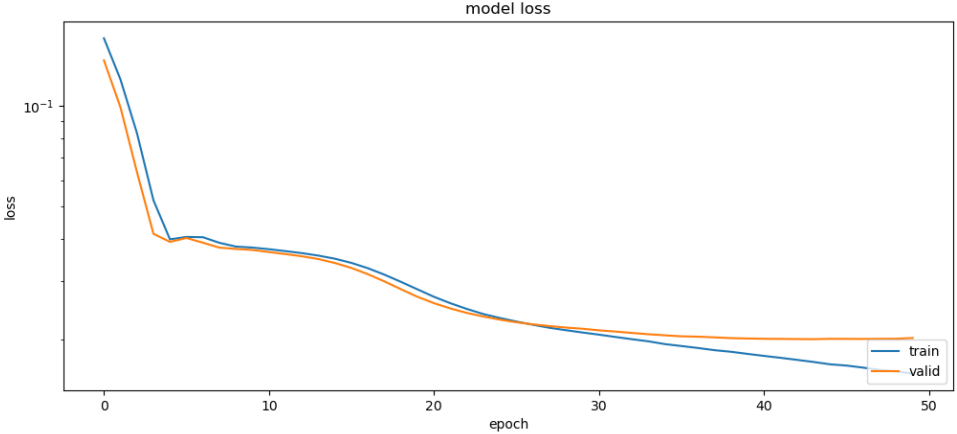

**Figure 3.** Training and test losses

**Codes:**

Code.zip contains tree models. The traditional models (MNB and LR) are in TraditionalModels.py, The best result is BestResultOnKaggleModel.py and the second one is SecondResultOnKaggleModel.py. Because I was thinking the second model is stronger in theory I wrote the report based on that.